\documentclass[fleqn,10pt]{wlscirep}
\usepackage[utf8]{inputenc}
\usepackage[T1]{fontenc}
\usepackage{chemformula} 
\usepackage[T1]{fontenc} 

\usepackage[]{graphicx}
\usepackage{epstopdf}

\title{Imaging and Identification of Single Nanoplastic Particles and Agglomerates}

\author[1]{Ambika Shorny}
\author[1]{Fritz Steiner}
\author[1]{Helmut Hörner}
\author[1,*]{Sarah M. Skoff}

\affil[1]{Atominstitut, Technische Universit\"at Wien, Stadionallee 2, Vienna, 1020, Austria}
\affil[*]{sarah.skoff@tuwien.ac.at}

\begin{abstract}
Pollution by nanoplastic is a growing environmental and health concern. Currently the extent of nanoplastic in the environment can only be cumbersomely and indirectly estimated but not  measured. To be able to quantify the extent of the problem, detection methods that can identify nanoplastic particles that are smaller than 1 $\mu$m are critically needed. Here, we employ surface-enhanced Raman scattering (SERS) to image and identify single nanoplastic particles down to 100 nm in size. We can differentiate between single particles and agglomerates and our method allows an improvement in detection speed of $10^{7}$ compared to state-of-the art surface-enhanced Raman imaging. Being able to resolve single particles allows to measure the SERS enhancement factor on individual nanoplastic particles instead of averaging over a concentration without spatial information. Our results thus contribute to the better understanding and employment of SERS for nanoplastic detection and present an important step for the development of future sensors.

\end{abstract}

\begin{document}

\flushbottom
\maketitle


The global increase in plastic production and disposal has resulted in large amounts of plastic  that end up in our environment. This was even further intensified by the recent COVID19 pandemic which required the widespread increased use of single-use-plastics for testing and personal protection equipment \cite{adyel_accumulation_2020}. 
Fragmentation of this plastic waste in the environment leads to micro- and nanoplastics which are dispersed even more easily and are harder to contain. Nanoplastics, commonly defined as particles with one dimension being smaller than 1 $\mu$m \cite{cai_analysis_2021}, are in particular very difficult to detect because of their small size but can cause great harm as these particles are able to cross the blood-brain barrier in living organisms and even enter the placenta \cite{campanale_detailed_2020}.
\begin{figure}[b!]
	\centering
	\includegraphics[width=0.8\linewidth]{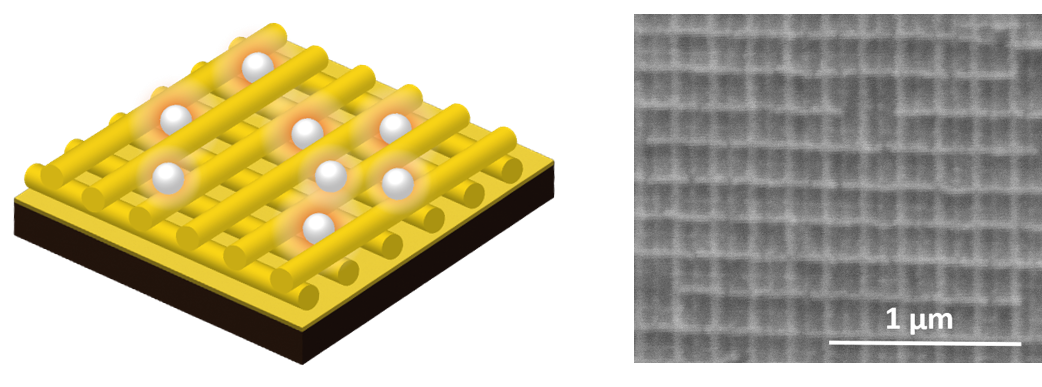}
    \caption{Sketch of SERS substrate with polystyrene beads (left) and scanning electron microscope image of SERS substrate used for the detection of nanoplastic (right).} 
	\label{fig:EFSERSGraph}
\end{figure}
Harmful effects on human health that arise due to continued exposure to micro- and especially nanoplastic are therefore an active field of research \cite{gruber_waste_2022, liu_bioeffects_2022}. 

As a first step in tackling the nanoplastic problem, suitable detection methods are required that can visualize and quantify the problem at hand. Thus far, the larger microplastic is mostly detected by visual inspection,  Fourier transform infrared (FTIR) spectroscopy \cite{kappler_analysis_2016}  and Raman microscopy \cite{hale_global_2020}. While visual inspection \cite{ivleva_microplastic_2017} is fastest and most widely used, it depends heavily on the interpretation of the observer and  also lacks chemical information. Further, it cannot be applied for nanoparticles. FTIR  and Raman spectroscopy are able to provide information on the vibrational structure of the detected molecules but still have some disadvantages. FTIR is restricted to larger particle sizes on the order of 20 $\mu$m and suffers from strong background signals from water which requires samples to be carefully dried. Raman microscopy works for smaller particles and can be used for spectroscopy in water but Raman scattering is an inherently weak process, with only about one in 10$^8$ photons of the excitation beam contributing to the Raman signal \cite{gardiner_practical_1989}. Other methods such as thermal analysis yields chemical information but is destructive and information about number and size is not provided \cite{shim_identification_2017} and scanning electron microscopy (SEM) can provide information on the geometry of small particles but requires careful sample preparation and makes it difficult to obtain chemical information. For these reasons, methods are required that offer good signal levels to ensure fast detection times, while still yielding information on the type of the analyte. 

Here, we show that by employing surface-enhanced Raman scattering (SERS), the Raman signals can be vastly enhanced by more than three orders of magnitude and by combining SERS with confocal microscopy, we are able to image and identify single plastic particles down to sizes of 100 nm.  While  surface-enhanced Raman spectroscopy has shown great benefits in the detection of small molecular coverages, where higher enhancement factors can be obtained as molecules can sample the smallest gaps of nanostructures \cite{kim_synthesis_2020,demirtas_facile_2020,zhang_large-scale_2016}, it is only very recently that it has been applied for the detection of nanoparticles and aerosols \cite{fu_surface-enhanced_2017,sivaprakasam_surface-enhanced_2021,yoo_novel_2022,alitahir_klarite_2020} and in particular nanoplastics \cite{chang_nanowell-enhanced_2022,fang_identification_2020,hu_quantitative_2022,jeon_detection_2021,kihara_detecting_2022,le_nanostructured_2021,lv_situ_2020,xu_surface-enhanced_2020,zhou_identification_2021,yang_identification_2022}. With our system, we are able to image and identify single nanoplastic particles and obtain the SERS enhancement factor without the need to average over different concentrations. Additionally, we demonstrate that we can decrease the image acquisition time for spatially-resolved nanoplastic detection by filtering the strongest Raman band instead of using Raman mapping and simultaneously imaging the particles by recording the scattered laser light, which we demonstrate to result in an improvement by a factor of $~10^7$ for spatially-resolved nanoplastic detection. Our detection method is not limited to any particular nanoplastic concentration and thus provides an important step towards the development of nanoplastic sensors in the future. 

\section{Materials and Methods}
\subsection{SERS substrate}
SERS enhancement is generally implemented in two different ways: either the analyte is mixed with a solution of plasmonic nanoparticles and then deposited on a non-SERS substrate \cite{lin_detecting_2022,lv_situ_2020,hu_quantitative_2022,kihara_detecting_2022,zhou_identification_2021}  or a plasmonic substrate is manufactured beforehand and the analyte is then deposited on it \cite{xu_surface-enhanced_2020,yang_identification_2022,jeon_detection_2021,le_nanostructured_2021,chang_nanowell-enhanced_2022}. The latter case has the advantage that in principle a higher reproduceability of the results should be possible. We therefore follow the latter method by employing a nano-patterned gold substrate (Pico Foundry \cite{noauthor_pico_nodate}) which consists of 3D crossed gold nanowires that form a square grid with a side length of 100 nm (Fig. \ref{fig:EFSERSGraph}). In our case, the enhancement of the Raman signal is due to electromagnetic enhancement. This means that as gold has a plasmon resonance in the visible, localized surface plasmons get excited and thus light gets tightly confined by the gold nanostructure forming "hot spots". The scattering of particles that sample these "hot spots" then also gets intensified by the plasmonic enhancement. 

\begin{figure}[t!]
	\centering
	\includegraphics[width=1\linewidth]{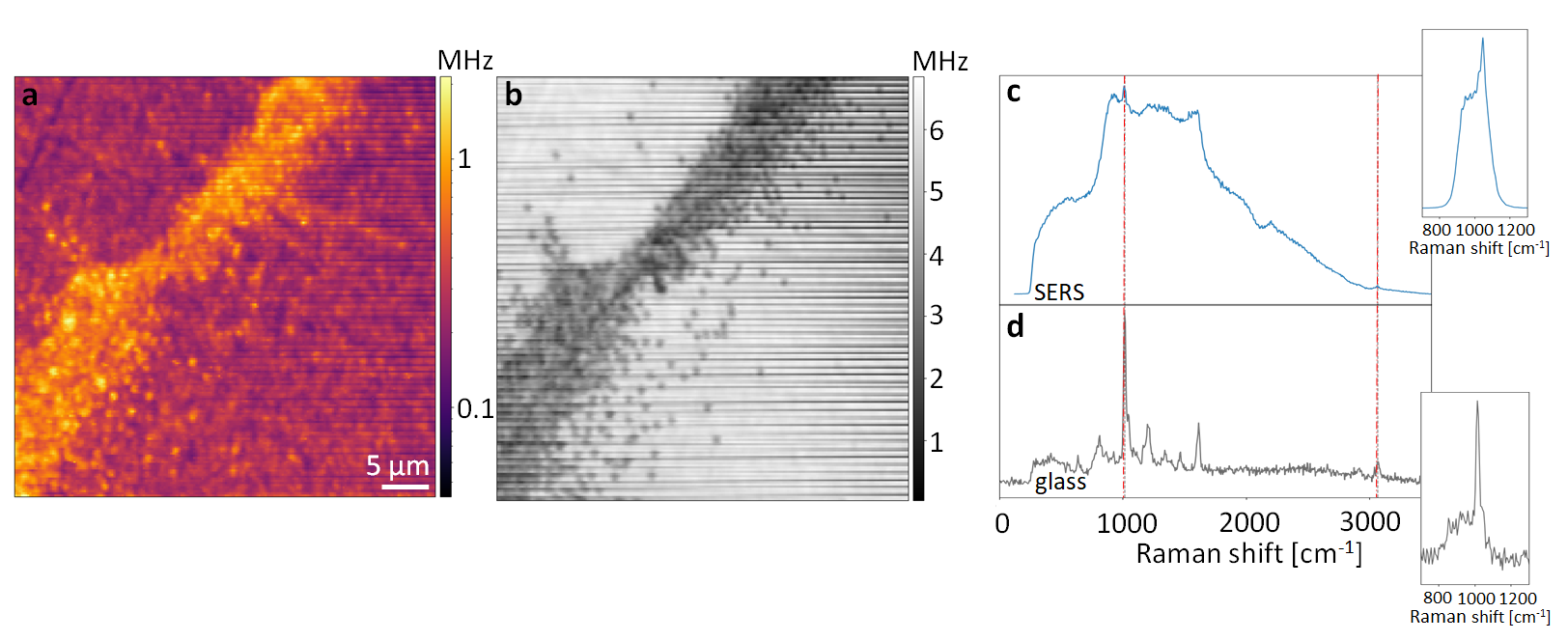}
    \caption{Agglomerates of and single 300 nm PS particles on the SERS substrate visualised (a) via surface-enhanced Raman imaging and (b) via recording the backscattered laser beam. (c) Raman spectra obtained from a single nanoparticle on a SERS substrate and on a  (d) reference glass substrate. Two unbroadened Raman modes that are visible in both spectra are indicated via dashed lines. The insets shows the respective Raman spectra with the bandpass filter to select the brightest C-C stretch vibrational mode.} 
	\label{fig:300nm}
\end{figure}
\subsection{Experimental Procedures}
The nanoplastic used was purchased from Sigma Aldrich in three different sizes: 800 nm, 300 nm and 100 nm diameter spherical polystyrene (PS) particles. They come dispersed in water (10\%) and before depositing them on the different substrates we dilute them further (1:1300) with ultrapure water to be able to also obtain single nanoparticles on the substrates.  
For detection of the nanoplastic on either substrate, a homebuilt confocal microscope is used. A 500 mW  laser at 638 nm is connected to the microscope setup via a fiber and the excitation beam is focussed via a Mitutoyo 50x objective with an NA of 0.55 onto the sample, which can be scanned by a stepper motor and a piezo in closed- or open-loop operation. The maximum power at the sample is about 24 mW.  The signal is  collected by the same objective and transmitted through a dichroic mirror and 75 $\mu$m pinhole. It is split into two beams by a 50/50 beamsplitter, which are filtered and then coupled into two optical fibers which lead to two single photon counting modules (SPCM). One of the beams is attenuated by neutral density filters before entering the detector, while for the other beam, we filter out the excitation beam completely by two longpass filters (LP 650 nm) and pick out the brightest part of the Raman band of polystyrene by a bandpass filter (BP 680 nm).

The backscattered light field consists of a contribution which is reflected by the substrate and a contribution that is affected by any particle present on the substrate by being either scattered or absorbed. When this light field is imaged as a function of position via the confocal microscope, the particles thus appear as shadows (Fig. \ref{fig:300nm} (b)), whereas in the other arm, where the brightest Raman mode is selected those particles that are plastic are bright on a dark background (Fig. \ref{fig:300nm} (a)). The maximum scan range of the stepper motor than can move the sample is 25 x 25 mm and that of the piezo is $100\times100$ $\mu$m. Typical pixel sizes of a piezo scan are between 100-300 nm. The typical time taken for closed-loop operation is 0.2 s per pixel, where 0.1 s integration time is used and the rest of the time is taken by the program to stabilize the piezo at the required position. To get an overview of a big area, the pixel size can be increased and the piezo can be operated without feedback with a lower integration time of down to 10 ms/pixel. Concerning our observed Raman scattering rate of the particles on the SERS substrate, integration times of less than 1 ms/pixel are possible and thus at the moment we are only limited by the way our piezo stage is implemented in our experimental control.

To obtain a wide Raman spectrum from about 290-4000 cm$^{-1}$, the bandpass filter is removed and the Raman emission can be sent to a spectrometer (Shamrock SR-303i, Andor Technology). We have used two different gratings of 300 l/mm and 600 l/mm and  typical integration times are between 1-10 s.
  \begin{figure}[t!]
	\centering
	\includegraphics[width=0.9\linewidth]{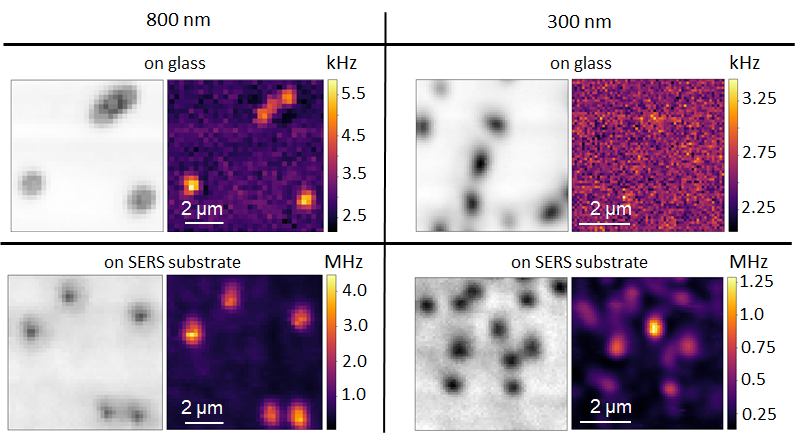}
    \caption{Comparison of confocal images taken of 800 nm and 300 nm polystyrene particles on a glass and a SERS substrate.} 
	\label{fig:GlassSERSComparison}
  \end{figure}

\section{Results and Discussion}
An example of an image of 300 nm polystyrene beads is displayed in Fig. \ref{fig:300nm} (a) and (b), where polystyrene beads with a diameter of 300 nm are imaged on the SERS substrate over an area of $40\times40$ $\mu$m. Sending the Raman scattered light at the position of the particles onto a spectrometer yields the known spectrum of polystyrene. Fig. \ref{fig:300nm}  shows the obtained spectrum of an individual particle on a (c) SERS substrate and (d) a reference glass substrate. When the particles are placed on the SERS target some Raman modes get broadened \cite{anema_surface-enhanced_2010} (Fig. \ref{fig:300nm} (c))  but one C-C stretch mode ($\nu_{24}$) and one C-H stretch mode ($\nu_{7}$) remain unaffected and thus can still be used for identifying the particles as polystyrene \cite{sears_raman_1981}. 

To obtain fast Raman images, we forego Raman mapping by filtering the strongest Raman mode at 1003 cm$^{-1}$ (C-C stretch mode) by a bandpass filter and sending this light to an SPCM to create the images. The spectrum with the bandpass filter introduced is displayed in the insets of Fig. \ref{fig:300nm} (c) and (d). The surface-enhanced Raman image displayed in Fig. \ref{fig:300nm} (a) shows agglomerates and individual 300 nm PS particles. Here, a log scale is applied to the colorbar to best visualise single particles as well as the brighter agglomerate. This stresses the importance of also being able to image individual nanoparticles to make sure the detection method is really suitable for the desired size range and down to the level of single particle detection.
To compare the results obtained from the SERS substrate to the Raman images obtained on a glass substrate, Fig. \ref{fig:GlassSERSComparison} shows confocal images taken of 800 nm and 300 nm polystyrene particles on both types of substrates. For 800 nm particles, both conventional Raman and SERS can resolve the particles, albeit SERS at a better signal to noise ratio. Importantly, however, for the  300 nm particles, the conventional Raman signal is already very small and thus the particles are barely visible, whereas on the SERS substrate they can be clearly distinguished from the background (Fig. \ref{fig:GlassSERSComparison}). This clearly shows that our here introduced SERS platform enables Raman detection of small nanoplastic particles in a facile fashion. By directly using appropriate filters instead of employing Raman mapping, these surface-enhanced Raman images can also be obtained at a much greater speed and integration times of $\le 1$ms/pixel are feasible (Supplementary Information and Figure S2). 
\begin{figure}[t!]
	\centering
	\includegraphics[width=1\linewidth]{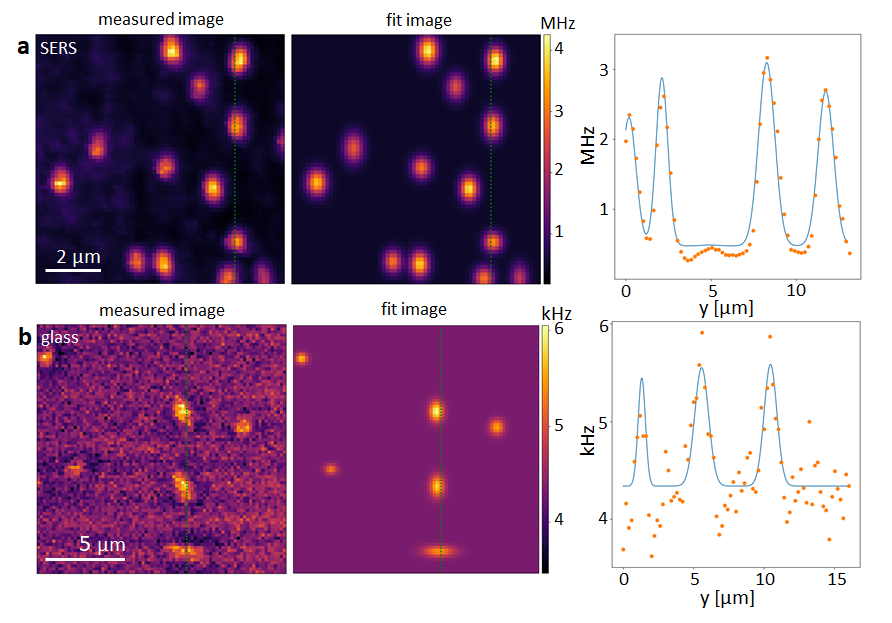}
    \caption{Measured Raman images and their corresponding fitted images for (a) the SERS substrate and (b) a glass substrate for 800 nm PS particles. On the right of these images a 2D slice at the position indicated in the respective images is plotted to further visualise the gain in absolute signal and in signal to background ratio when imaging these nanoparticles via SERS.} 
	\label{fig:contrast800}
\end{figure}
To evaluate the experimental enhancement factor of the SERS substrate and quantify the improvement of the images in terms of contrast, we fit the images to the point spread function of the scatterers (Fig. \ref{fig:contrast800}), which is well approximated by a two-dimensional Gaussian \cite{qu_nanometer-localized_2004}. This yields the peak intensity of the Raman emission and we calculate the experimental enhancement factor as: 
\begin{align}
    \text{EF} = \frac{\text{I$_{\text{SERS}}$}}{\text{I$_{\text{glass}}$}},
    \label{eq:enhancement}
\end{align}
where I$_\text{SERS}$ is the peak intensity of the particles on the SERS substrate and I$_\text{glass}$ is the peak intensity on the glass substrate for the same Raman band. In both cases the background has been substracted.
\begin{figure}[t!]
	\centering	
 \includegraphics[width=0.9\linewidth]{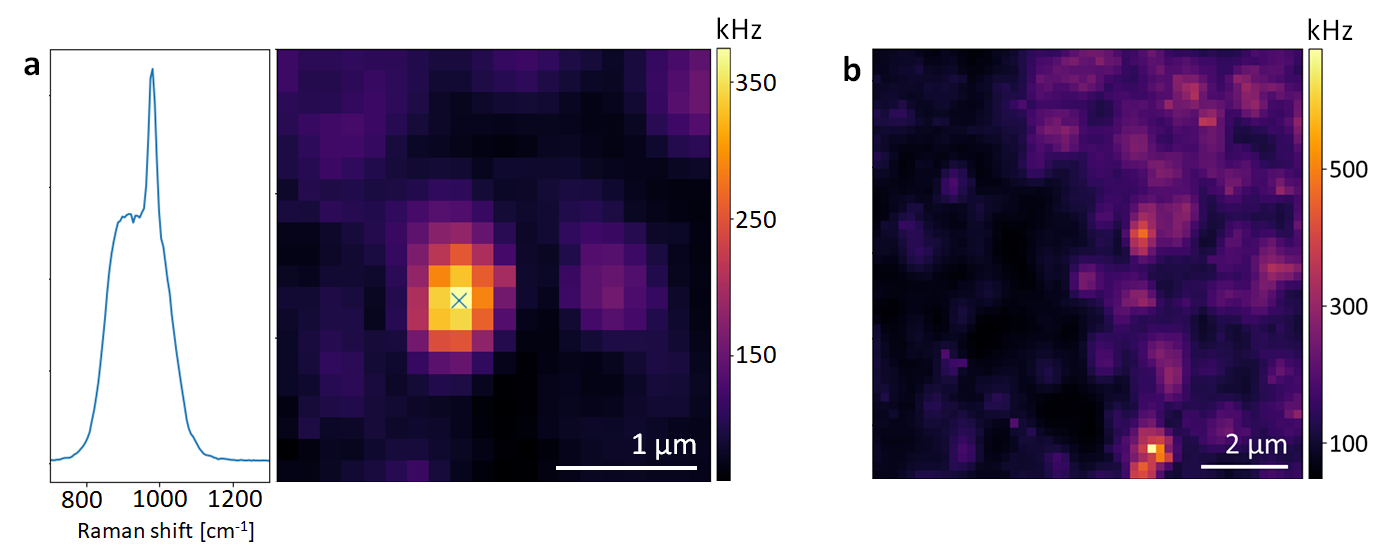}
    \caption{(a) Spectrum and surface-enhanced Raman image of a 100 nm PS particle. (b) Raman image of agglomerates and single 100 nm nanoplastic particles} 
	\label{fig:100nm_sers}
\end{figure}
We find that the SERS substrate enhances the Raman signal by more than 3 orders of magnitude (Table \ref{table:1}). 
\begin{table}[b!]
\centering
\begin{tabular}{ |c|c|c|c| } 
 \hline
 particle size & S/B glass & S/B SERS substrate & EF \\ 
 \hline
 800 & 1.849 $\pm$0.098 & 6.425 $\pm$ 0.839 & 979 $\pm$189 \\ 
 \hline
 300 & 1.116 $\pm$ 0.033 & 3.743 $\pm$ 1.165&2219 $\pm$ 1109 \\ 
 \hline
  100&-&2.205$\pm$ 0.327&-\\
  \hline
\end{tabular}
\caption{Experimental enhancement factor (EF) and the signal to background ratio (S/B) for single nanoparticles of different sizes on a glass or SERS substrate.}
\label{table:1}
\end{table}
Such a high enhancement factor has not been measured yet on single nanoplastic particles \cite{xu_surface-enhanced_2020} and our results show that being able to image single particles is key to obtaining the true enhancement factor for detection of particles of such small sizes. Employing SERS, we also increase the signal to background ratio by more than an order of magnitude.  This is visualised in Fig. \ref{fig:contrast800}, where a slice through an image is taken, depicting the particles as peaks in this 2D plot. The values for the enhancement factor  and  the signal to background ratio are given in table \ref{table:1} for a laser power of $23.32 \pm 0.85$ mW  at the target, which was the highest power used and the best in terms of contrast for the glass substrate. While we are only able to see Raman images of the plastic particles on glass for excitation laser powers of $>$ 4 mW, for the SERS substrate powers as low as 150 $\mu$W show a clear signal.
For lower powers the experimentally evaluated enhancement factor therefore represents a lower limit. In terms of signal to background ratio, there is a tendency that even for the SERS substrate this ratio decreases as the size of the particles decreases. This is not surprising as the size of the particles gets smaller and thus the fraction of light from our diffraction limited excitation laser beam of about 1.45 $\mu$m  that illuminates the particles with respect to the illuminated area of the background, decreases as well.
Employing SERS for nanoparticles is generally a very new field that has only recently shown promising results \cite{lin_detecting_2022,lv_situ_2020,hu_quantitative_2022,kihara_detecting_2022,zhou_identification_2021,xu_surface-enhanced_2020,yang_identification_2022,jeon_detection_2021,le_nanostructured_2021,chang_nanowell-enhanced_2022}, but previously, enhancement factors for SERS have been determined for different concentrations of the analyte without any spatial information, which leaves a large uncertainty on the number of particles that actually contribute to the signal \cite{hu_quantitative_2022}.  Often, the advantage of SERS is illustrated by stating a minimum detectable concentration of nanoplastic \cite{le_nanostructured_2021, kihara_detecting_2022, zhou_identification_2021}, which is still larger than the concentrations of hundreds to thousands of ng/l that are expected in the environment \cite{li_impacts_2023,enders_abundance_2015,pabortsava_high_2020}  and also lacks spatial information. Being able to image single particles is thus an important step towards developing precise sensors in the future that are able to detect low concentrations of single nanoplastic particles rather than only agglomerates and for understanding and optimizing SERS microscopy for nanoparticles in general.
Due to the enhancement in Raman signal obtained from the SERS substrate we are able to detect and image individual particles with a diameter of only 100 nm,  which is far below the diffraction limited spot of our laser beam. A SERS image of a 100 nm particle with the corresponding spectrum with bandpass filter can be seen in Fig. \ref{fig:100nm_sers}(a) and a larger image of agglomerates of and single 100 nm particles is seen in Fig. \ref{fig:100nm_sers}(b). This clearly shows that the plasmonic enhancement allows for imaging nanoplastic particles that otherwise remain invisible.

\section{Conclusion}
In a time where not much is known yet about the distribution of nanoplastic particles in the environment \cite{lins_nanoplastic_2022}, techniques need to be developed to quantify and monitor this pollution. This is crucial as it is expected that nanoplastic is at least as widespread as microplastics, while having the potential to be more harmful to living organisms \cite{lehner_emergence_2019,liu_cellular_2021}.

In this study, we have demonstrated the imaging and identification of single nanoplastic particles and of agglomerates down to sizes of 100 nm via surface-enhanced Raman microscopy with a gain in acquisition speed of 10$^7$ compared to state-of-the art micro-Raman measurements \cite{le_nanostructured_2021,kihara_detecting_2022,hu_quantitative_2022, zhou_identification_2021}.
 We also demonstrate the variation in signal strength between single particles and agglomerates and hence the need for spatial information to assess the detection limits of the developed methods. For particles of 300 nm and 800 nm, we compare our measurements to Raman microscopy on a non-SERS substrate and find enhancement factors of more than three orders of magnitude. To our knowledge, these are the largest enhancement factors seen so far for individually imaged nanoplastic particles \cite{xu_surface-enhanced_2020} and are needed to image particles of such a small size in such a facile way.  

Instead of using Raman mapping we employ filtering of the brightest Raman band of the analyte to obain direct Raman images using our confocal microscope and simultaneously image the particles by recording the backscattered laser light. Compared to Raman mapping which usually takes 1-10 s integration time per pixel, this is a vast improvement \cite{chang_nanowell-enhanced_2022} as integration times of less than 1 ms/pixel become feasible for an image of individual nanoplastic particles. Our way of imaging can be even further improved by employing more narrowband filters \cite{sun_tunable_2022} possibly also for different Raman bands. In addition, if one wanted to record the whole spectrum for a wide wavelength range, spectra only need to be taken for those coordinates where bright particles are visible in the confocal images. Currently, simultaneous imaging of the backscattered laser light lets us image particles down to sizes of 300 nm, but this resolution can be further improved by better control over the phase of the laser beam \cite{taylor_interferometric_2019}. While not being particle specific, this way of imaging also allows a quick overview of interesting parts of the sample with minimal laser powers.

Nanoplastic concentrations in the environment span a wide range \cite{li_impacts_2023,enders_abundance_2015,pabortsava_high_2020} and hence it is crucial to develop sensitive methods that also work on the single particle level. Our results not only aid in understanding and employing SERS enhancement for nanoplastic detection but are an important step for the process of developing the first in-situ detector to monitor nanoplastics in the environment in the future.

\section{Data availability}
The datasets used and/or analysed during the current study are available from the corresponding author on reasonable request.
\section{Acknowledgments}
We want to thank A. Rauschenbeutel for the loan of equipment.
This work was supported by the Austrian Research Promotion Agency (FFG) (no. 884447, PhoQus2D). S.M.S. acknowledges funding from the Austrian Science Fund (FWF) (no. V934-N, Quantoom).

\section{Authors' contributions}
A.S. performed all the measurements and analysed the data. F.S. analysed the data and H.H. set up the experimental control. S.M.S. conceived the idea, supervised the project and wrote the manuscript with input from all the authors.

\section{Competing interests}
The authors declare no competing interests.

\bibliography{references,detectionNanoplasticBib,SERSnanoplasticpaper,SERSnanoplasticpaper2}
\end{document}